\documentclass[a4paper,11pt]{article}
\usepackage{pos}

\title{Confinement/Deconfinement in 4D compact QED on the lattice}

\author*[a,b]{Lee C. Loveridge}
\author[b]{Orlando Oliveira}
\author[b]{Paulo J. Silva}

\affiliation[a]{Los Angeles Pierce College,\\
  6201 Winnetka Ave, Woodland Hills, USA}

\affiliation[b]{CFisUC, Department of Physics,University of Coimbra,\\  Coimbra, Portugal}

\emailAdd{loverilc@piercecollege.edu}
\emailAdd{orlando@uc.pt}
\emailAdd{psilva@uc.pt}

\abstract{It has long been known that there is a phase transition between confined and unconfined phases of compact pure gauge QED on the lattice. In this work we report three manifestations of this phase change as seen in the Landau gauge photon propagator, the static potential, and distribution of Dirac Strings in the gauge fixed configurations. Each of these was calculated with large lattices with volumes: $32^4$, $48^4$ and $96^4$. We show that the confined phase manifests with a Yukawa type propagator with a dynamically generated mass gap, a linearly increasing potential, and a significant concentration of Dirac strings while the unconfined phase appears consistent with the continuum results: a free propagator, a near constant long-distance potential, and a small concentration of Dirac strings trending towards zero. Furthermore, the photon propagator is investigated in detail near the transition between the two phases.}

\FullConference{%
 The 38th International Symposium on Lattice Field Theory, LATTICE2021
  26th-30th July, 2021
  Zoom/Gather@Massachusetts Institute of Technology
}

\begin{document}
\maketitle

\section{Introduction}

The first major question is why we should study $U(1)$ or QED on the lattice. After all, QED is fairly well understood in the free theory limit and interactions can be described well with perturbation theory. However, $U(1)$ gauge theories are relevant to understand the Higgs sector and many condensed matter problems. QED corrections to lattice QCD problems are becoming increasingly important, and $U(1)$ gauge theories are a laboratory for modeling gauge theories on quantum computers. The compact formulation of $U(1)$ gauge theory on a lattice shows two different phases that are associated with different values of the coupling constant. Our particular interest here is to understand the "photon" propagator and the static potential in both a confining and a deconfining phase. 

\section{Setup and Action}

A compact lattice $U(1)$ gauge theory is set up in a very direct way. As with lattice QCD, the action is the sum of the real part of plaquettes which in the continuum limit become the standard Maxwell tensor composed of electric and magnetic fields. The action is
\begin{equation}
       S_W (U) = \beta \sum_x \sum_{1 \leqslant \mu, \nu \leqslant 4} \left\{ 1 - \Re \, \left[ U_{\mu\nu} (x) \right]\right\} \,, \, \beta = 1/ e^2
   \label{Eq:accao}
\end{equation}
where the plaquette $U_{\mu\nu}(x)$ is given by 
\begin{equation}
   U_{\mu\nu} (x) = U_\mu (x) \, U_\nu (x + a \, \hat{e}_\mu) \, U^\dagger_\mu (x + a \, \hat{e}_\nu) \,  U^\dagger_\nu (x)
   \label{Eq:plaquette}
\end{equation}
and the links $U_{\mu}(x)$ are defined as 
\begin{equation}
   U_\mu (x) = \exp\left\{ i \, e \, a\, A_\mu \left( x + \frac{a}{2} \, \hat{e}_\mu \right) \right\}.
   \label{Eq:Link}
\end{equation}
We analyzed this theory using standard hybrid monte-carlo sampling.

\section{Static Potential}
\begin{figure}
\begin{center}
\includegraphics[width=0.6\textwidth]{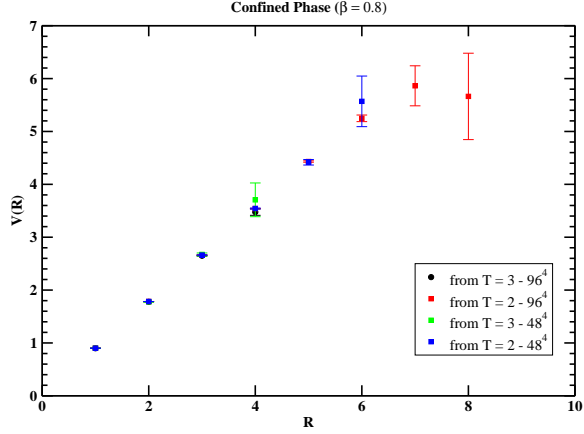}
\end{center}
\caption{\label{fig:conf-pot} Static potential in the confined $\beta=0.8$ state.}
\end{figure}

\begin{figure}
\begin{center}
\includegraphics[width=0.6\textwidth]{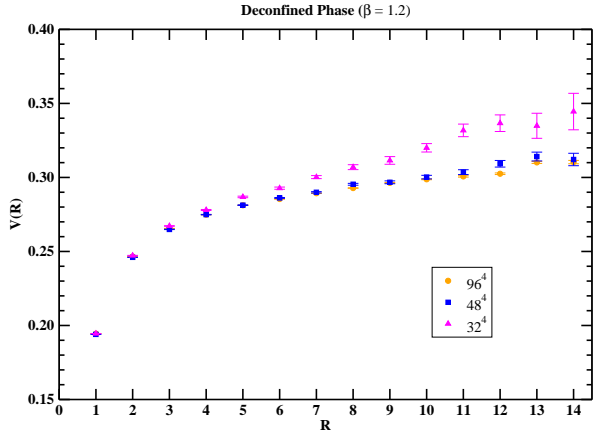}
\caption{\label{fig:deconf-pot} Static potential in the deconfined $\beta=1.2$ state.}
\end{center}
\end{figure}

Our first interesting result is in the static potential. Figures \ref{fig:conf-pot} and \ref{fig:deconf-pot} show a clear phase change in the static potential between $\beta=0.8$ ($e^2=1.25$) and $\beta=1.2$ ($e^2=0.83$). In the low $\beta$ (large coupling) regime, we see a potential that is rising linearly with distance, while at high $\beta$ (low coupling) we see a potential that grows much more slowly with distance.

\begin{figure}
\begin{center}
    \includegraphics[width=0.6\textwidth]{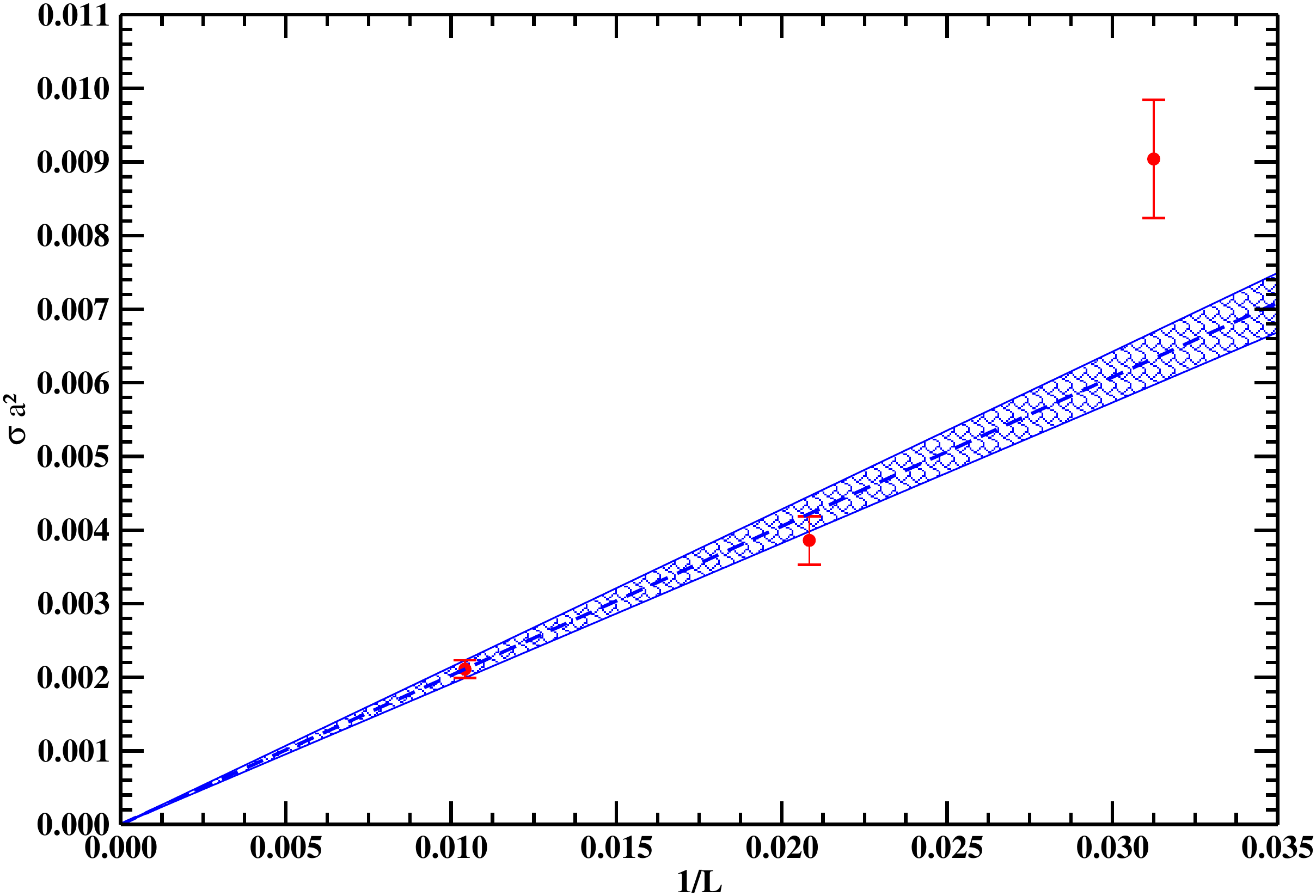}
\end{center}
\caption{\label{fig:string tension} String tension in the infinite volume limit.}
\end{figure}

The strength of the linear part of the potential is characterized by a string tension $\sigma$. In figure \ref{fig:string tension} we plot the low coupling string tension of the various lattice sizes from figure \ref{fig:deconf-pot}. The string tension appears to  approach zero as the size of the lattice increases. This suggests that the infinite volume continuum limit may be a free field theory.

\section{Photon Propagator in Landau Gauge}

The photon propagator is given by the expectation value of photon field between two different momentum states:
\begin{equation}\langle A_\mu (p_1) ~ A_\mu(p_2) \rangle = V \, \delta( p_1 + p_2 ) \, D_{\mu\nu}(p_1).
\end{equation}
Before averaging the field values, we perform gauge shifts so that the propagator will be in the Landau Gauge and take the following form:
\begin{equation}
D_{\mu\nu}(p) = \left( \delta_{\mu\nu} - \frac{p_\mu p_\nu}{p^2} \right) \, D(p).
\end{equation}

There is some ambiguity in how we should recover the photon $(A_\mu)$ field from the lattice links  
\begin{equation}
  U_\mu (x) = \exp\left\{ i \, e \, a\, A_\mu \left( x + \frac{a}{2} \, \hat{e}_\mu \, \right)\right\}.  
\end{equation}

The simplest solution is what one might call the linear definition and follows the approach of non-abelian gauge theories. We assume that the argument of the exponential is small, that our link $U_{\mu}$ is nearly 1, i.e. the argument $eaA_{\mu}$ is nearly 0, and thus we can find the photon field by taking the imaginary part of the configuration
\begin{equation}
 A_\mu \left( x + \frac{a}{2} \hat{e}_\mu \right)  = \frac{U_\mu(x) - U^\dagger_\mu(x)}{2 \, i}.  
 \label{eq:lin-def}
\end{equation}
This method is commonly used when studying $SU(3)$ (QCD) on the lattice where there is a physical scale and one can be certain that the argument of the exponent is small and the links are indeed close to unity. Since we are exploring $U(1)$ with different coupling constants, we have no physical scale and therefore cannot be sure that the argument is small.

We will instead use a logarithmic definition of $A_\mu$.
\begin{equation}
A_\mu \left( x + \frac{a}{2} \hat{e}_\mu \right) = -i \, \ln \Big(  U_\mu (x) \Big)
 \label{eq:log-def}
\end{equation}
This definition is exact up to machine accuracy, but it is much harder as it requires taking a logarithm. However, logarithms of $U(1)$'s pure phase configurations are much simpler to take than logarithms of $SU(3)$'s matrix configurations.

The two different definitions of the photon field require different gauge fixing formulations to ensure orthogonality. For the linear definition (\ref{eq:lin-def}), we maximize the functional 
\begin{equation} 
F[U; g] = \frac{1}{V \, D} \sum_{x,\mu} \, \Re \left[  \,  g(x) \, U_\mu (x) \, g^\dagger(x + a \, \hat{e}_\mu )\,  \right],
\label{eq:lin-func}
\end{equation}
while for the logarithmic definition (\ref{eq:log-def}) we must maximize the functional
\begin{equation}
\widetilde{F}[U;g] = \frac{1 }{V \, D } \sum_{x ,\mu} \,  \bigg\{ 1 -  a^2 e^2 \left[ A^{(g)} _\mu \left( x + \frac{a}{2} \hat{e}_\mu \right) \right]^2 \bigg\}.
\label{eq:log-func}
\end{equation}
We achieved the best results by first maximizing the linear functional (\ref{eq:lin-func}) and then maximizing the logarithmic functional (\ref{eq:log-func}).

\begin{figure}
\begin{center}
\includegraphics[width=0.6\textwidth]{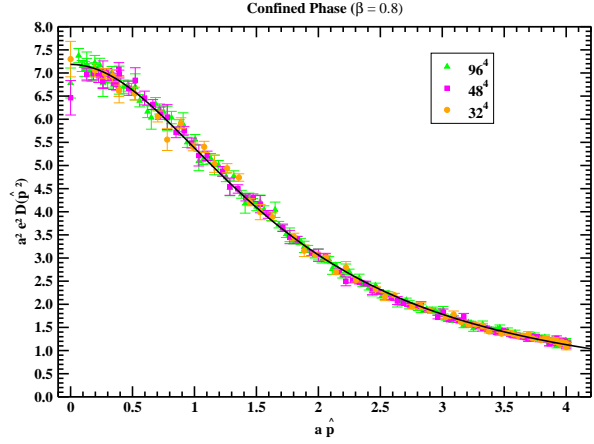}
\end{center}
\caption{\label{fig:conf-prop}Confined $\beta=0.8$ propagator.}
\end{figure}

After gauge fixing and computing the photon propagator, we see that in the confined $\beta=0.8$ phase (figure \ref{fig:conf-prop}), the propagator takes the standard form of a propagator with well defined mass:
\begin{equation} 
a^2 e^2 D(a^2\hat{p}^2) = \frac{z_0}{(a\hat{p})^2 + (am)^2}.
\end{equation}

\begin{figure}
\begin{center}
\includegraphics[width=0.6\textwidth]{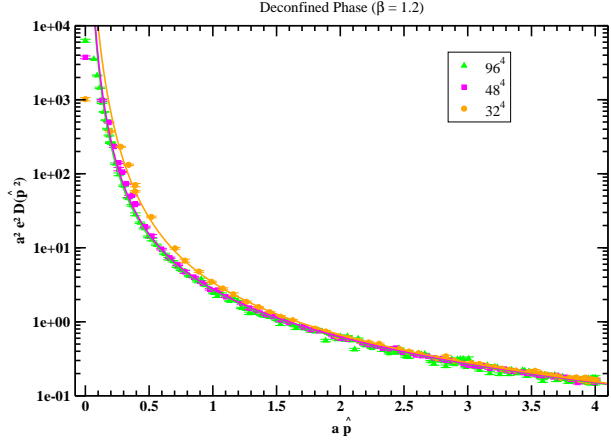}
\caption{\label{fig:deconf-prop}Deconfined $\beta=1.2$ propagator.}
\end{center}
\end{figure}

By contrast, in the deconfined $\beta=1.2$ phase (figure \ref{fig:deconf-prop}), the propagator appears to be approaching that of a massless free photon: 
\begin{equation}
    e^2a^2  D(a^2 \hat{p}^2) = \frac{Z_0}{( a \,  \, \hat{p})^2 } + \frac{Z_1}{( a \,  \, \hat{p})^4 }.
\label{eq:deconf-prop}
\end{equation}
The coefficient $Z_1$ is effectively a measure of how different our calculated propagator is from a true massless propagator. If we compare the values of $Z_0$ and $Z_1$ as the lattice size increases, it is clear that $Z_0$ is fairly stable while $Z_1$ approaches 0, suggesting that it is an artifact of the finite size and spacing of the lattice.

\begin{figure}
\begin{center}
\includegraphics[width=0.6\textwidth]{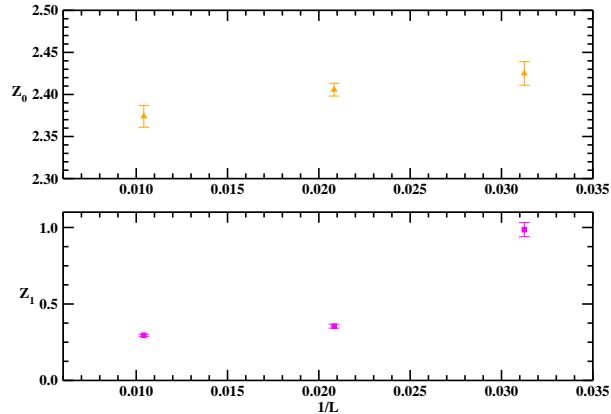}
\caption{\label{fig:Z0Z1} Coefficients $Z_0$ and $Z_1$ from equation \ref{eq:deconf-prop} for different lattice sizes. Note that $Z_0$ is nearly constant while $Z_1$ approaches 0 as the lattice size increases.}
\end{center}    
\end{figure}

\section{Causes and Nature of the Transition}
We do not yet understand the cause of the phase transition. It may be a result of Dirac Strings in the configurations. This is when the links of a plaquette make a full $2\pi$ rotation so that the log of the plaquette differs from the total of the logs of the legs by a multiple $m_{\mu\nu}$ of 2 $\pi$:
\begin{equation} 
U_{\mu\nu} (x) = \exp\left\{ i \, e \, a \, \Big( \, \Delta A_{\mu\nu} (x) \, \Big)  \right\};
\end{equation}
\begin{equation}
   \Delta A_{\mu\nu} (x)  =  \sum_\text{loop} A_\mu
+ \frac{ 2  \pi  m_{\mu\nu} (x)}{e  a }.
\end{equation}

The total of all such Dirac Strings appears to be gauge invariant, but they do move from one plaquette to another or from one site to another in the gauge fixing process. We report here the non-gauge invariant average number of strings per site:
$$ m = \frac{1}{6 \, V} \, \sum_{x, \mu <\nu}  | m_{\mu\nu}(x) |.$$ This variable $m$ is analogous to the magnetization observed in condensed matter via the Ising model.

\begin{figure}
\begin{center}
\includegraphics[width=0.6\textwidth]{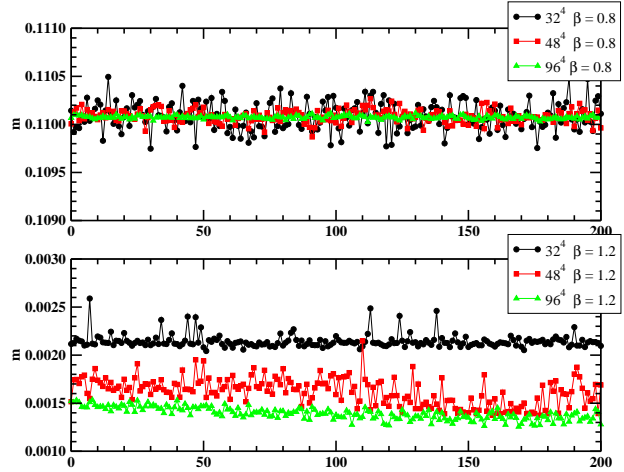}
\caption{\label{fig:monopoles} Dirac Strings in the confined and deconfined phases.}
\end{center}
\end{figure}
Note from figure \ref{fig:monopoles} that the average number of Dirac Strings is more than an order of magnitude higher for the confined phase than for the deconfined phase. Furthermore, the average number of strings appears to be independent of lattice size in the confined case but appears to be decreasing with increased lattice size in the deconfined phase. This suggests that Dirac Strings may play a role in the confined phase while being absent in the infinite size and continuum limit of the deconfined phase.

\begin{figure}
    \includegraphics[width=\textwidth]{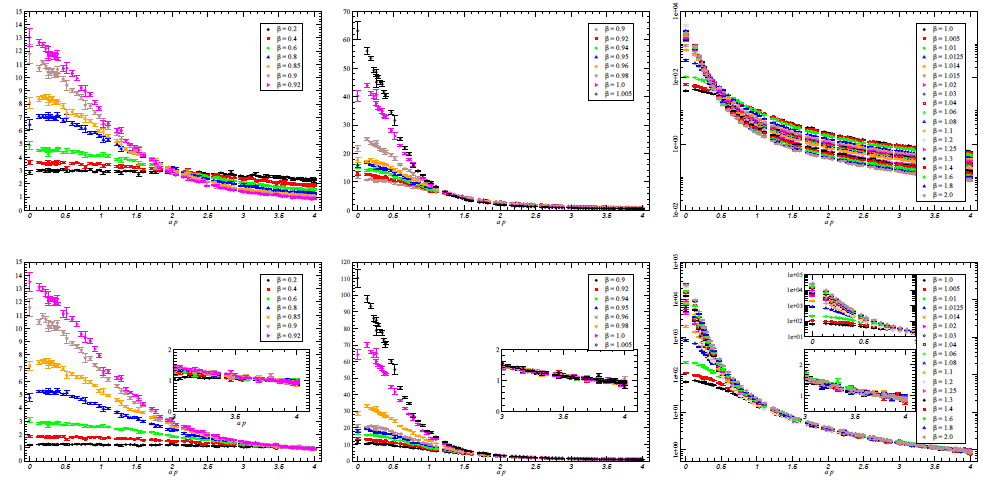}
\caption{\label{fig:prop-trans} Photon propagators at various values of beta. The upper row is raw data from the simulation while the bottom row has renormalized all data to have consistent behavior at large values of momentum.}
\end{figure}

A closer look at propagators for various values of $\beta$ suggests a first order transition between the phases. Figure \ref{fig:prop-trans} shows photon propagators at various values of $\beta$. The upper row is the raw data from the simulation while the lower row has been normalized so that all propagators have similar high momentum behavior. It is clear that all of the low $\beta$ propagators show a massive behavior and values under 10 at low momentum. By contrast, the large $\beta$ propagators have very large values at low momentum, suggesting that they may be free propagators in the infinite volume limit. 

\begin{figure}
\begin{center}
    \includegraphics[width=0.66\textwidth]{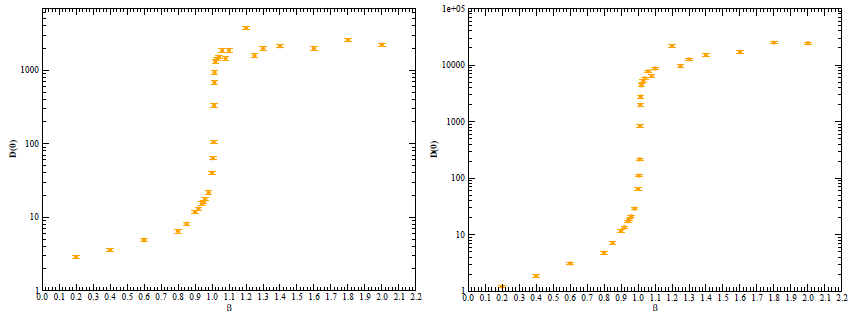}
\caption{\label{fig:D0-trans} The zero momentum component of the photon propagator plotted versus $\beta$ for both the raw and normalized data.}
\end{center}
\end{figure}

Figure \ref{fig:D0-trans} shows just the zero momentum component of the propagator plotted versus $\beta$ for the raw (left) and normalized (right) data. It is clear that this value changes by several orders of magnitude in the neighborhood of $\beta=1.0$; note in particular, the near vertical slope at $\beta=1.0125$. This suggests a first order transition between the confined and deconfined phases near $\beta=1.0$ with the sharpest change at $\beta=1.0125$. More details can be found in \cite{Loveridge21.1} \cite{Loveridge21.2} and references therein.

\section*{Acknowledgements}

This work was partly supported by the FCT –Fundação para a Ciência e a Tecnologia, I.P., under Projects Nos. UIDB/04564/2020 and UIDP/04564/2020. P. J. S. acknowledges financial support from FCT (Portugal) under Contract No. CEECIND/00488/2017. The authors acknowledge the Laboratory for Advanced Computing at the University of Coimbra (http://www.uc.pt/lca) for providing access to the HPC resource Navigator.

\end{document}